\documentclass[twocolumn]{aastex7}

\newcommand{\htt}{H$_2$}
\newcommand{\water}{H$_2$O}

\usepackage{array}
\usepackage{hyperref}
\usepackage{longtable}
\usepackage{rotating, graphicx}
\usepackage{xcolor}
\newcolumntype{P}[1]{>{\centering\arraybackslash}p{#1}}
\newcolumntype{R}[1]{>{\raggedleft\arraybackslash}p{#1}}
\newcolumntype{L}[1]{>{\raggedright\arraybackslash}p{#1}}

\usepackage{lineno}

\received{July 9, 2025}
\revised{September 5, 2025}
\accepted{September 5, 2025}
\submitjournal{ApJ}

\shorttitle{}
\shortauthors{Gibbs et al.}

\begin{document}\linenumbers

\title{Simulations of Flare Chemistry in Brown Dwarf Companions to Active M Dwarfs}



\author[0000-0002-9027-4456]{Aidan Gibbs}\email{gibbs.aidan@gmail.com}
\affiliation{Department of Physics \& Astronomy, University of California, Los Angeles, CA 90095, USA}

\author[0000-0002-0176-8973]{Michael P. Fitzgerald}\email{mpfitz@ucla.edu}
\affiliation{Department of Physics \& Astronomy, University of California, Los Angeles, CA 90095, USA}

\begin{abstract}


Brown dwarfs that are short period ($<10\,$day) companions to actively flaring M dwarfs may provide a context to directly observe flare-driven photochemistry and structural changes in an extrasolar planet-like atmosphere. To assess the viability of directly observing flare impacts in the atmosphere of a brown dwarf, we perform self-consistent temperature-chemistry modeling of the atmospheric response to individual energetic superflares. We modified the existing open-source \texttt{VULCAN} chemical-kinetics and \texttt{HELIOS} radiative-transfer codes for this purpose. Similar to previous studies of flare impacts on hydrogen dominated atmospheres, we find flares are capable of orders-of-magnitude changes in the mixing abundances of many chemical species, including important opacity sources like CH$_4$ and CO$_2$. However, due to fast chemical timescales resulting from high temperatures and densities in brown dwarf atmospheres, these changes last for a short-period of time, generally less than a day, and are only plausibly observable via high resolution emission spectroscopy. We find that the most observable, short-term spectral changes in hot (T$_{\text{eff}}\sim2000\,$K), high-gravity ($\log{\text{g}}\sim5$), cloudless brown dwarfs are the photolysis of \water\ and enhancement of CO$_2$, which can result in part-per-thousands spectral changes in the hours after a flare.


\end{abstract}

\keywords{Brown dwarfs, Exoplanets, Planetary atmospheres, Stellar flares}


\section{Introduction} \label{sec:intro}

Recent surveys have shown that the occurrence rate of small, roughly Earth-sized planets around M dwarfs is higher than that of other stellar types (e.g. \citealt{Mulders2015,Bashi2020,Sabotta2021,Pinamonti2022,Ment2023}). In addition to having a higher frequency of small planets orbiting them, M dwarfs have small radii and low luminosity that can make characterizing the planets around them relatively easier than characterizing planets around earlier stellar types (e.g. \citealt{Kaltenegger2009_transmission_terrestrial_planets, Triaud2021_smallstaropportunity}). 

While the frequency of small planets increases around later stellar types, so too does the frequency of stellar activity (e.g. \citealt{Pietras2022}). M dwarfs in particular can be active on gigayear timescales \citep{Kiman2021}, much longer than more massive stars. Because M dwarfs can maintain stellar activity on long time scales alongside otherwise potentially habitable planets, the effect of stellar activity on planets around M dwarfs has become an important area of investigation. Recent investigations have shown that flares can have important chemical and thermodynamic impacts on model atmospheres of terrestrial planets \citep{Segura2010,Tilley2019,Louca2023,Ridgway2022}. While the effects of stellar activity on terrestrial planets have begun to be explored in simulations, no observational effects are likely to be observed in the near-term due to the high inherent challenge of detection of secondary atmospheres on small, Earth-sized planets, which is complicated further by the presence of stellar activity through phenomena like the Transit Light Source Effect \citep{Rackham2018}.

In contrast, hot Jupiters and brown dwarfs have atmospheres that are relatively easier to characterize than terrestrial atmospheres, but are rare \citep{Kiefer2019}, especially around M dwarfs \citep{Sabotta2021,Gan2022}. Because they are rare around flare stars, and because it is unknown if the atmospheres of Jovian planets and brown dwarfs can form and support life (see discussions in \citealt{Sagan1976_Jovian_life,Yates2017_Y_dwarf_life,Lingam2019_brown_dwarf_life}), the impacts of flares and other stellar activity on hot Jupiters and brown dwarfs is not as well explored theoretically as in smaller planets. Investigations of flare impacts on hot Jupiter atmospheres have only begun recently by \cite{Hazra2021,Louca2023, Konings2022}, and \cite{Nicholls2023}.

While brown dwarfs on short-period orbits around M dwarfs are rare, a number of systems are known and more are currently being discovered by studies using the \textit{Transiting Exoplanet Survey Satellite (TESS)} and ground-based transit surveys \citep{Vowell2025}. \textit{TESS} light curves of transiting brown dwarfs also conveniently provide some characterization of the stellar activity level of host stars through the detection of optical flares and rotational spot modulation. Using \textit{TESS}, we are beginning to discover some brown dwarfs on short-period orbits around actively flaring M dwarf hosts (e.g. \citealt{Gillen2017_AD3116,Irwin2018_LP26175C,Jackman2019_NGTS7Ab,Carmichael2022_TOI2119}). In these systems, stellar activity may be partly induced by tidal locking between the star and brown dwarf, which helps maintain a fast stellar rotation rate \citep{Kotorashvili2024_Mdwarf_spinup}.

These new discoveries could provide opportunities to observationally characterize the impacts of stellar flares, or stellar particle radiation from coronal mass ejections (CMEs) and proton events, on a planet-like atmosphere outside the solar system. However, no simulations have yet been done to predict what the effects of flares or other radiation will look like on the atmospheres of the high-gravity environments of brown dwarfs. The objective of this paper, therefore, is to extend the work of \cite{Louca2023, Konings2022}, and \cite{Nicholls2023} to explore flare induced photochemistry in higher gravity and higher temperature atmospheres and the observability of these effects, for the first time. Like \citet{Nicholls2023}, we will also explore temperature-chemistry coupling and feedback in these atmospheres.

In Section \ref{sec:model}, we describe our computational framework to model the photochemistry and thermodynamic impact of flares on a brown dwarf atmosphere. Section \ref{sec:res} presents the changes in atmospheric chemistry and temperature-pressure structure resulting from individual flares of varying duration and energy. We compare these results to previous work and discuss observability in Section \ref{sec:diss} with a summary and future outlook in Section \ref{sec:conc}.

\section{Methods} \label{sec:model}

\begin{figure*}[t]
\centering
\includegraphics[width=1.0\textwidth]{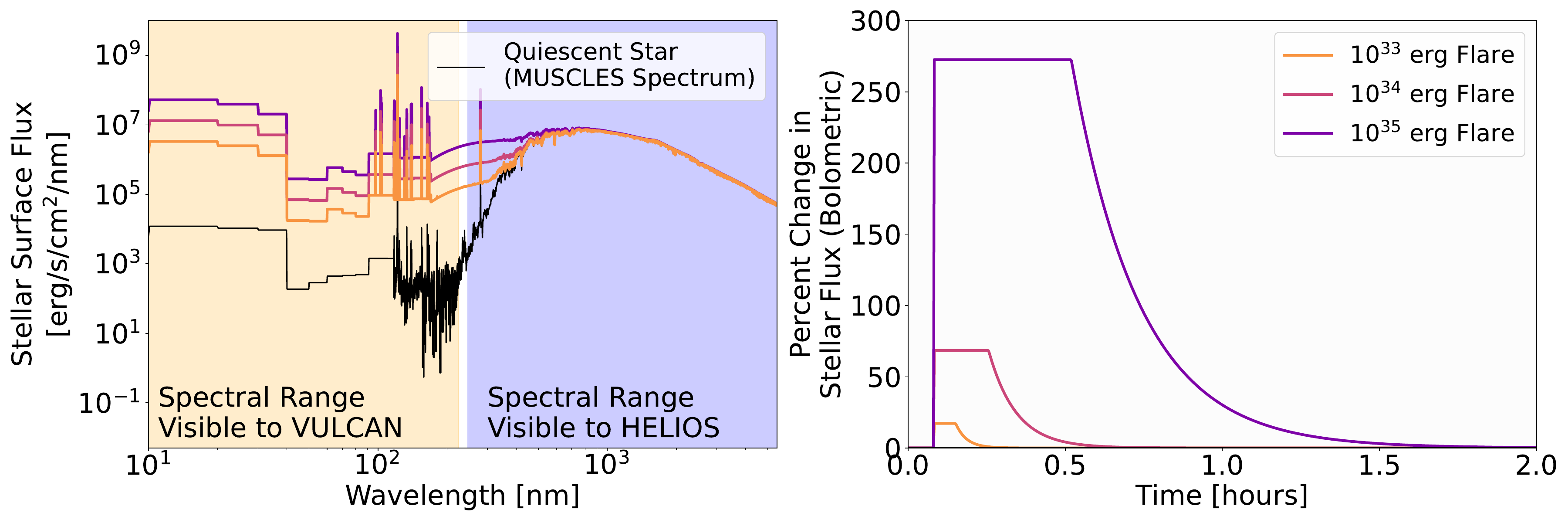}
\caption{\textbf{Spectral and temporal evolution of \texttt{Fiducial Flare} synthetic flares.} \textbf{Left:} The quiescent stellar spectrum (MUSCLES GJ 832, \citealt{France2016}) compared to the star plus flare spectrum at peak luminosity for flares of three different energies. Radiation longer than $\sim240\,$nm is not capable of photodissociation and is effectively invisible to \texttt{VULCAN}, while radiation shorter than $\sim240\,$nm is invisible to \texttt{HELIOS} due to limitations in opacities. \textbf{Right:} The bolometric luminosity light curve for flares of different energies relative to the quiescent stellar bolometric luminosity.}
\label{fig:flare_profiles}
\end{figure*}

\subsection{Model Framework}\label{sec:framework}

To model the photochemical and thermodynamic impact of M dwarf flares on brown dwarf atmospheres, we adapt existing open-source software. We simulate atmospheric chemistry with the \texttt{VULCAN} 1-D photochemical kinetics code \citep{Tsai2021,Tsai2017}, which is coupled to \texttt{HELIOS} \citep{Malik2017,Malik2019}, a 1-D radiative-convective transfer code for determining the atmospheric pressure-temperature profile. Atmospheric temperature and chemistry are strongly interconnected, with chemistry determining atmospheric opacities, and temperature controlling chemical reaction rates. Together, these codes provide a self-consistent temperature-chemistry model. To model the time-dependent spectrum of a flare as input for \texttt{VULCAN} and \texttt{HELIOS}, we use the \texttt{Fiducial Flare} package (\citealt{Loyd2018}; \href{https://github.com/parkus/fiducial_flare}{https://github.com/parkus/fiducial\_flare}). In the following subsections, we will provide a high-level overview of each package's function and modifications we have made. Further information can be found in the original publication for each package.

\subsubsection{\texttt{VULCAN} 1-D Photochemical Kinetics Model}\label{sec:vulcan}

\texttt{VULCAN} solves the 1-D continuity equation for major chemical species in an atmosphere given a set of elements to include, the initial abundance of each of those elements in the atmosphere, a temperature-pressure profile for the atmosphere, and an incoming top-of-atmosphere stellar flux. For a given set of elements, it includes a dictionary of thermochemical and photochemical reactions, the chemical network, which can act to produce or destroy each chemical species. Chemical species can also be transported between atmospheric layers by molecular and thermal diffusion, eddy diffusion, and user specified vertical winds. Benchmark tests of \texttt{VULCAN} compared to other models can be found in \cite{Tsai2017} and \citep{Tsai2021}.

Typically, \texttt{VULCAN} assumes the incoming stellar flux and temperature-pressure profile is constant. We have modified this in a similar fashion to \cite{Louca2023} to allow a time-dependent stellar flux that changes due to stellar flares, as well as allowing the temperature-pressure profile to change. This is accomplished by prompting \texttt{VULCAN} to update these data between time step iterations as necessary, which propagates into changes in chemical reaction rates. 

\texttt{VULCAN} is capable of using a number of different chemical networks of neutral chemical species, including custom user networks. Chemistry between charged species is not yet publicly available, an important limitation that will be discussed further in Section \ref{sec:diss}. The simplest chemical network includes only H, C, and O elements, while the most complex network that has been publicly tested includes H, C, N, O, S, Ti, and V. Increasing the numbers of elements included in the network significantly increases runtime, as each new element will have reactions with all other elements. The atmospheres of hot brown dwarfs have metal species like TiO and metal hydrides as important opacity sources in their atmospheres. For this reason, we opt to use the most complex published network, which includes HCNOSTiV reactions, for all of our simulations. In heavily irradiated brown dwarfs, ionized metal species and H$-$ can also become important opacity sources, but are necessarily excluded. While this does not replicate the full chemical complexity of a hot brown dwarfs atmosphere, it will reproduce the major chemical behavior of the most abundant species, and provide clues as to the behavior of other metal species.

Technically, the photochemical network of \texttt{VULCAN} is only tabulated between $500$ and $2500\,$K. However, in our model atmosphere, the highest pressures (above $\sim10,$bar) have temperatures exceeding $2500\,$K, reaching nearly $6000\,$K at $10^3\,$bar. Within this pressure and temperature region, equilibrium chemistry would be expected to dominate even in the presence of vertical mixing and photochemistry, and the chemical profiles computed by \texttt{VULCAN} closely match the profiles produced by the equilibrium chemistry code \texttt{FastChem}, which has been validated for temperatures up to $6000\,$K \citep{Kitzmann2018_FastChem,Kitzmann2022_FastChem2}. 


\texttt{VULCAN} has the option to set boundary conditions for atmospheric escape from the top and bottom of the atmosphere. As we are simulating high-gravity brown dwarfs, we assume there is no significant escape from the atmosphere and use \texttt{VULCAN}'s default zero flux boundary condition for both the top and bottom of the atmosphere, meaning mass and elemental composition is conserved. \texttt{VULCAN} photoreaction cross-sections are limited to $\sim10\,$nm at the shortest wavelength, meaning that VULCAN is not capable of simulating X-ray or more energetic photochemistry, which can produce secondary photoionization and dissociation cascades \citep{Locci_2022_xray_photochem}. In addition, no reactions in \texttt{VULCAN} include enthalpies of reaction, meaning that endothermic and exothermic reactions have no impact on the thermodynamics of the atmosphere. Photon energy absorbed in the photodissociation of a species is assumed to be fully consumed in dissociation without changing the temperature of the reaction products. 

\subsubsection{HELIOS 1-D Radiative-Convective Model}\label{sec:helios}

\texttt{HELIOS} solves the radiative transfer equation in the 1-D plane parallel, two stream approximation. For our purposes, it's most important feature is the ability to calculate the layer-by-layer atmospheric opacity `on-the-fly' given an atmospheric chemical mixing ratios. This means that the chemistry calculated by \texttt{VULCAN} can be fed to \texttt{HELIOS} and used to recalculate atmospheric opacity.

For all our \texttt{HELIOS} runs, we use absorption opacities of \water, $\text{CO}_2$, CO, $\text{CH}_4$, $\text{NH}_3$, HCN, $\text{C}_2\text{H}_2$, $\text{H}_2\text{S}$,  TiO, and VO, and scattering opacities of $\text{H}_2$, He, \water, $\text{CO}_2$, and CO. We additionally use collision induced absorption opacities of $\text{H}_2$--$\text{H}_2$ and $\text{H}_2$-He.

We have modified \texttt{HELIOS} so that atmospheric opacity can be updated between iterations as necessary, without restarting a new run or the time-consuming process of reloading opacity files. Like \texttt{VULCAN}, we have also modified it to allow updates to the stellar flux between time step iterations in response to a flare.

\texttt{HELIOS} allows an optional convective readjustment step at the end of radiative transfer calculations to check for and simulate convectively unstable layers. We turn this option on for all simulations, with the default criteria $\kappa=2/7$ for an ideal diatomic gas. However, given that convection only operates in high pressure regions of the atmosphere where photochemistry is not occurring, we expect it will only potentially be important for determining the quiescent, pre-flare state of the atmosphere and will have no effect on the time-dependent impact of the flares. In practice, we observe that the convective readjustment step never triggers during our simulations and our atmosphere is fully radiative within our selected pressure boundaries.

\texttt{HELIOS} inherently assumes local thermodynamic equilibrium (LTE) in its simulations, which is valid for dense regions for the atmosphere where collisions are frequent, but is not valid for low density regions with infrequent collisions. The impact of non-LTE effects on the upper atmospheres in response to flares will be left to future work.

\subsubsection{Fiducial Flare} \label{sec:fidflare}

Like \cite{Louca2023} and \cite{Konings2022}, we use the \texttt{Fiducial Flare} python package \citep{Loyd2018} to generate synthetic time-dependent stellar flare spectra. \texttt{Fiducial Flare} produces flares with a simple light curve that has the form of a boxcar followed by exponential decay. Flares of arbitrary energy can be produced by changing the flare equivalent duration, with more energetic flares lasting longer and with higher peak luminosities. The spectral energy distribution of generated flares is a $9000\,$K blackbody with the addition of some important UV lines, such as Ly-$\alpha$.  Examples of the light curve and spectra of several flares with varying energies is shown in Figure \ref{fig:flare_profiles}.

While we assume this synthetic flare formulation reasonably approximates flares with the bulk of their energy in the optical/UV, there is important variability in the spectral and temporal nature of real flares (i.e. optical energy cannot always be trusted to predict UV energy or duration, for example \citealt{Brasseur2023_uvoptical_flares,Namekata2017_solartype_flares}). We do not attempt to capture this complexity and flares with significantly different spectral energy distributions (e.g. predominately X-ray) could produce different results. 

Another point of note is that many stellar flares are expected to be accompanied by particle events such as Coronal Mass Ejections (CMEs) and proton events. Due to observational difficulty, however, very few CMEs have been observed from a star other than the Sun (e.g. \citealt{Argiroffi2019_extrasolar_CME}), and there is still significant uncertainty as to whether M dwarfs can produce CMEs and/or proton events, and how precisely they may differ from those of the Sun \citep{Vida2019_Mdwarf_CMEs,Wood2021_Mdwarf_winds}. Previous work has shown that high energy particle radiation can have a significant impact on planet-like atmospheres, often more significant that the impact of electromagnetic radiation \citep{Airapetian2016_YoungSunEarthimpactofCMEs,Tilley2019,Hazra2021}. Even if M dwarfs do not typically produce CMEs with flares, short period brown dwarfs could still experience significant particle fluxes due to their proximity and potential for interconnected magnetospheres. With our current computational framework, we cannot model the impact of particle radiation, and leave this to future work. It is also possible that direct magnetic heating could occur in the atmosphere due to the interaction with and movement of the brown dwarf through the stellar magnetic field (e.g. \citealt{Cohen2024_magneticheating_trappist}), which will not be explored in this work.  

\subsection{Coupling \texttt{VULCAN} and \texttt{HELIOS}} \label{sec:coupling}


\subsubsection{Initial Quiescent Atmosphere} \label{sec:qscnt_atmo}

To create the initial quiescent atmosphere, we begin by running \texttt{VULCAN}, which requires an input stellar spectrum, initial abundance of elements, and temperature-pressure profile. For all simulations we use a MUSCLES spectrum \citep{France2016} of the M1.5V dwarf GJ 832 as the quiescent stellar spectrum of the host star. MUSCLES spectra cover X-ray to sub-millimeter wavelengths and are a combination of observed and synthetic spectra. The abundance of all elements in our simulations is set to Solar. For the initial temperature-pressure profile, we use the profile of a Sonora-Bobcat \citep{Marley2021} model with the same effective temperature and surface gravity as our desired brown dwarf. We then run \texttt{VULCAN}, using the equilibrium chemistry code \texttt{FastChem} \citep{Kitzmann2018_FastChem} to determine the initial chemical profiles, which \texttt{VULCAN} iteratively updates to include the effects of vertical transport and photochemistry until it has reached a steady state.

Once \texttt{VULCAN} has reached a steady state chemistry, the final mixing profiles are used as input to \texttt{HELIOS} for calculating atmospheric opacity. \texttt{HELIOS} sees the same stellar spectrum as \texttt{VULCAN}, however, due to limitations in the resolution of opacities, it is binned and resampled to a spectral resolution of only R$=50$. \texttt{HELIOS} starts with an isothermal temperature-pressure profile determined from the estimated equilibrium temperature of the brown dwarf, and iterates until it reaches a new steady state profile based on the atmospheric opacity and internal and external radiation. As the temperature-pressure profile of our brown dwarf is dominated by the internal temperature, $T_{eff}\approx T_{internal}$. 

Once \texttt{HELIOS} finishes its initial run, \texttt{HELIOS} and \texttt{VULCAN} began running in an iterative leapfrog sequence. A new \texttt{VULCAN} run is initialized with the end chemistry of its last run (to reduce runtime) and the new temperature-pressure profile determined by \texttt{HELIOS}. Next \texttt{HELIOS} runs again. Because \texttt{HELIOS} runs are relatively quick compared to \texttt{VULCAN}, \texttt{HELIOS} starts every run from the same initial isothermal profile (the \texttt{HELIOS} default) rather than the profile from the last run, although with opacities updated based on \texttt{VULCAN} mixing chemistry. Both codes take turn running in this manner until a stable model atmosphere is found where temperature does not change by more than $1\,$K and chemical abundances do not change by more than $\sim1\%$ in any layer between runs. For our model brown dwarf, it takes approximately 15 back-and-forth iterations to achieve this stability. While this framework produces a stable, self-consistent, quiescent 1-D atmosphere, it should be noted that real brown dwarfs exposed to frequent stellar activity are likely constantly variable at some level, with no single stationary quiescent state. Even isolated brown dwarfs can be highly variable with temperature and chemistry changing at some level across longitude and latitude (e.g. \citealt{Biller2024_variability_wise1049,McCarthy2025_simp0136_variability}).

\subsubsection{Flare Impacted Atmosphere} \label{sec:flare_atmo}

To simulate the impact of a flare on the quiescent atmosphere, \texttt{VULCAN} and \texttt{HELIOS} again run in a sequential, leapfrog manner similar to the process used to generate the quiescent atmosphere. The major differences are that now each code sees a time-dependent stellar flux driven by the addition of a synthetic flare, and each code is only allowed to run for a short time step before sending its results to the other package. This timestep, $\Delta t_\mathrm{sequential}$, must be short enough such that the chemistry has not changed enough to significantly impact temperature and temperature has not changed enough to significantly impact chemistry, as well as being short enough to well-resolve the lightcurve of the flare. We determine $\Delta t_\mathrm{sequential}=10\,$s to be a good balance that meets these criteria while not being so short that there is excessive switching between codes, which is the limiting factor in simulation run time.

Within this $\Delta t_\mathrm{sequential}$, both \texttt{VULCAN} and \texttt{HELIOS} run on their own independently determined time steps. The time step for \texttt{VULCAN}, $\Delta t_\mathrm{VULCAN}$, is determined automatically based on reaction rates to ensure numerical stability, but we artificially limit it to be a maximum of $1\,$s so that \texttt{VULCAN} must take at least 10 steps in every run. \texttt{HELIOS} does not have adaptive time step capabilities for real time runs such as ours, meaning we must manually set the time step, $\Delta t_\mathrm{HELIOS}$, for each run, which should be on the order of the fastest radiative timescale in the atmosphere. The radiative timescale in our model atmosphere near the top of the atmosphere at $10^{-5}\,$bar is $\sim1\,$s. \texttt{HELIOS} time steps are limited to $1.0\,$s and must complete exactly 10 iterations for each run. As a test of computational stability, we also try a $10\,$s \texttt{HELIOS} time step and find no notable difference in results. 

The longest wavelength of photo-dissociation cross-sections included in \texttt{VULCAN} is $\sim240\,$nm. Conversely, the shortest wavelength for opacities included by default in \texttt{HELIOS} is $246.6\,$nm. This limit in \texttt{HELIOS} is based on the potential for more energetic radiation to cause ionization and dissociation, but also somewhat on the limited availability of absorption opacities at shorter wavelengths. This means that the redder portion of the stellar spectrum is essentially invisible to \texttt{VULCAN}, while the bluer portion is invisible to \texttt{HELIOS}. For photochemistry, this is not a significant concern since the longer wavelengths do not strongly impact chemistry, but for radiative transfer considerations, it means that $\sim50\%$ of the energy from our simulated flares, which occurs short of $246\,$nm, cannot go into heating the atmosphere, whether it is consumed by photochemistry or not. Consequently, our results will underestimate the bulk heating of the brown dwarf atmosphere due to a flare by potentially up to a factor of 2. However, because most chemical species typically have larger absorption cross-sections at shorter wavelengths, which causes that radiation to be absorbed at lower atmospheric densities and pressures than longer wavelengths, our results for heating in the lowest pressure regions ($<1\,$mbar) of our model atmosphere could be underestimated by a greater amount. The addition of non-LTE or particle effects can also enhance heating, meaning that our estimates for heating from flares is a definitive lower bound of heating from stellar activity.

Both codes are also capable of simulating flare impacts independently from each other, by omitting updates from the other code. This is useful for isolating the impact of chemistry on thermodynamics and vice versa, which is explored in Section \ref{sec:res}.

\section{Results} \label{sec:res}

\subsection{Properties of the Simulated System} \label{sec:sys_prop}

Our goal is model the photochemical and thermodynamic impact of flares on the atmospheres of brown dwarfs which have properties similar to known brown dwarfs that are close companions to active M dwarfs, which can be observed with current facilities. As such, the constructed stellar and brown dwarf properties of our simulated system are generally similar to known systems like TOI-2119 \citep{Carmichael2022_TOI2119}, LP 261-75 \citep{Irwin2018_LP26175C}, and NGTS-7A \citep{Jackman2019_NGTS7Ab}, but is not meant to be an exact replica of any of these systems. For reference, these systems all have brown dwarfs with masses $\gtrsim 65\,\text{M}_{\text{Jup}}$ on orbits of $<0.07\,$au around early or mid M dwarfs. We choose to model a brown dwarf with effective temperature $T_{\text{eff,BD}}=2000\,$K and surface gravity $\log{g} = 5.0$ (so that a brown dwarf with the same values is available in the Sonora model grid) orbiting a GJ 832 stellar analog with $R_\star=0.45\,R_\odot$ at a semi-major axis of $0.05\,$au with zero eccentricity. We adopt a radius of $1.08\,\text{R}_{\text{Jup}}$ (the same radius as TOI-2119 b), which corresponds to a mass of  $\sim 47 \,\text{M}_{\text{Jup}}$.   

The quiescent chemical mixing profiles of select species are shown as solid lines in the panels of Figure \ref{fig:1e34_2hour_chem}, and the quiescent temperature profile of the atmosphere is shown in the top left panel of Figure \ref{fig:temperature}. We simulate single flares with energies of $10^{33}, 10^{34}$ and $10^{35}\,$ erg bolometric energies. One of the largest flares and CMEs ever observed from the Sun, the Carrington event, which is frequently used as a reference point, is estimated to have had a total energy on the order of $\sim10^{32}\,$erg \citep{Cliver2013}. Solar events approaching this magnitude are rare, of similar frequency to the solar cycle, and impact Earth even less frequently, with the Carrington event possibly being the most energetic solar storm to directly impact Earth in the last $\sim200$ years. In comparison, highly active M dwarfs like AU Mic can produce flares with optical energies of $10^{33}-10^{34}\,$ as frequently as every few days to weeks \citep{Gilbert_2022}. One example system, TOI-2119 was observed to have $20$ flares with optical energies $\gtrsim10^{32}$\,erg in \textit{TESS} sectors 24 and 25, as estimated by \cite{Pietras2022}.  


\subsection{Impact of a Single $10^{34}\,$erg Flare}\label{sec:sfe}

Using the methodology described in Section \ref{sec:framework}, we simulate the chemical and thermal evolution of the brown dwarf atmosphere for two hours following the onset of a single $10^{34}\,$erg superflare. The simulation begins with five minutes of quiescent stellar flux to verify atmospheric stability before the flare initiates and the incoming flux changes. We also independently run a simulation of the same length without any flare to verify that atmospheric evolution is a result of the flares and not computational instability. In testing, we always observe that our model atmosphere trends back towards the original quiescent state after a perturbation in the stellar flux is applied and then removed, which further supports that the quiescent atmosphere is in a stable state. The total simulation time is limited by computational run time, with a two-hour simulation requiring approximately one day to run on a single GPU.

\begin{figure*}[t]
\centering
\includegraphics[width=1.0\textwidth]{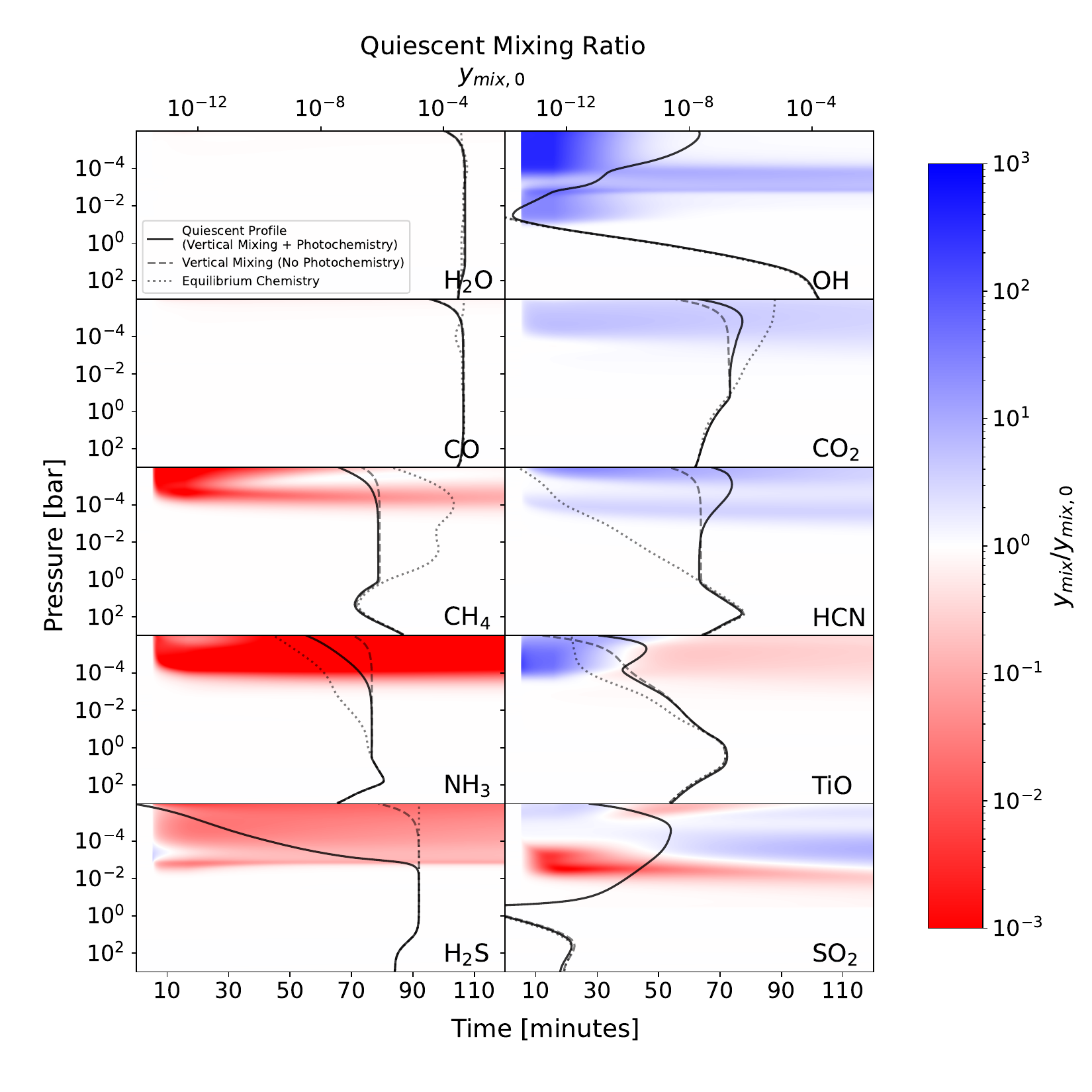}
\caption{\textbf{Temperature-coupled photochemistry in response to a single $10^{34}\,$erg flare.} The flare begins at five minutes. The initial (quiescent) atmospheric mixing profiles of each chemical species is shown with solid lines, and includes the effects of vertical mixing and photochemistry. We also show mixing profiles calculated for the same temperature profile, but without photochemistry (dashed), and without vertical mixing or photochemistry (i.e. equilibrium chemistry, dotted), so that the effect of both photochemistry and vertical mixing on the quiescent profiles can be seen. Red shading indicates a decrease in abundance compared to the initial mixing profile, while blue shading indicates an increase in abundance. The temperature-coupled simulation is limited to two hours due to computational cost.}
\label{fig:1e34_2hour_chem}
\end{figure*}

Changes in the mixing ratios of major atmospheric gases (excluding \htt\ and He) over the two-hour duration are presented in Figure \ref{fig:1e34_2hour_chem}. As expected, no significant variations occur during the initial five-minute quiescent phase. Once the flare begins, most atmospheric species exhibit order-of-magnitude or greater changes in their mixing ratios, both during the flare and its immediate aftermath. These changes predominantly occur in the upper atmosphere at pressures lower than 1 mbar, though some species experience alterations extending lower in altitude, to nearly the 0.1 bar pressure level. 

The greatest pressure level at which a given species experiences changes in mixing ratio is determined by a complex interplay of photochemistry and vertical mixing. In the absence of these processes, the mixing ratio of most chemical species will vary smoothly with temperature and pressure. The addition of vertical mixing, however, causes `quenching', where, at a certain pressure (the quench pressure), the mixing timescale becomes shorter than the chemical equilibrium timescale. At altitudes above this pressure, the equilibrium chemistry is too slow to adjust, and the species' mixing ratio remains nearly fixed at the quench pressure value. The addition of photochemistry further modifies this profile by producing or destroying species at low pressures, disrupting the uniform mixing profile expected from vertical mixing alone. The impact of photochemistry and vertical mixing can be seen in the three mixing profiles shown for each species in Figure \ref{fig:1e34_2hour_chem}. The dotted profiles show only equilibrium chemistry, the dashed profiles add vertical mixing (but not photochemistry), and the solid profiles are our quiescent chemistry which includes both photochemistry and vertical mixing. As an additional note, when vertical mixing is included, the quiescent mixing abundance of most species (for example \water\ and CO) shows a decrease at pressures below $10^{-5}\,$bar. This is not due to photochemistry, but instead marks roughly the location of the turbopause/homopause, where molecular diffusion becomes stronger than eddy diffusion at higher altitudes and the mixing abundances of heavier species decrease compared to lighter species.

For highly photoreactive species like OH and SO$_2$, photochemical production at low pressures leads to orders-of-magnitude abundance enhancements that can be transported downward to the quench pressure, creating ``pinch points'' where equilibrium and photochemistry mixing profiles intersect. Less photoreactive species, such as HCN and CH$_4$, may still experience photochemical changes at low pressures, but these alterations do not extend down to the quench pressure. Instead, these species develop regions of fixed mixing ratios at mid-altitudes, sandwiched between the equilibrium chemistry profile at high pressures and photochemistry-dominated behavior at lower pressures.

During a flare, the atmospheric depth at which photochemistry occurs for a given species remains largely unchanged. Instead, the reaction rates within those regions temporarily increase due to the enhanced stellar flux. This is because even in high-energy cases, flares are short-lived (typically lasting only a few hours at most) and do not significantly alter the optical depth of the atmosphere to ionizing and dissociating radiation. As shown in Figure \ref{fig:1e34_2hour_chem}, the impact of the flare is mostly confined to the same pressure levels that are already affected by photochemistry in the quiescent atmosphere.

\begin{figure*}[t]
\centering
\includegraphics[width=1.0\textwidth]{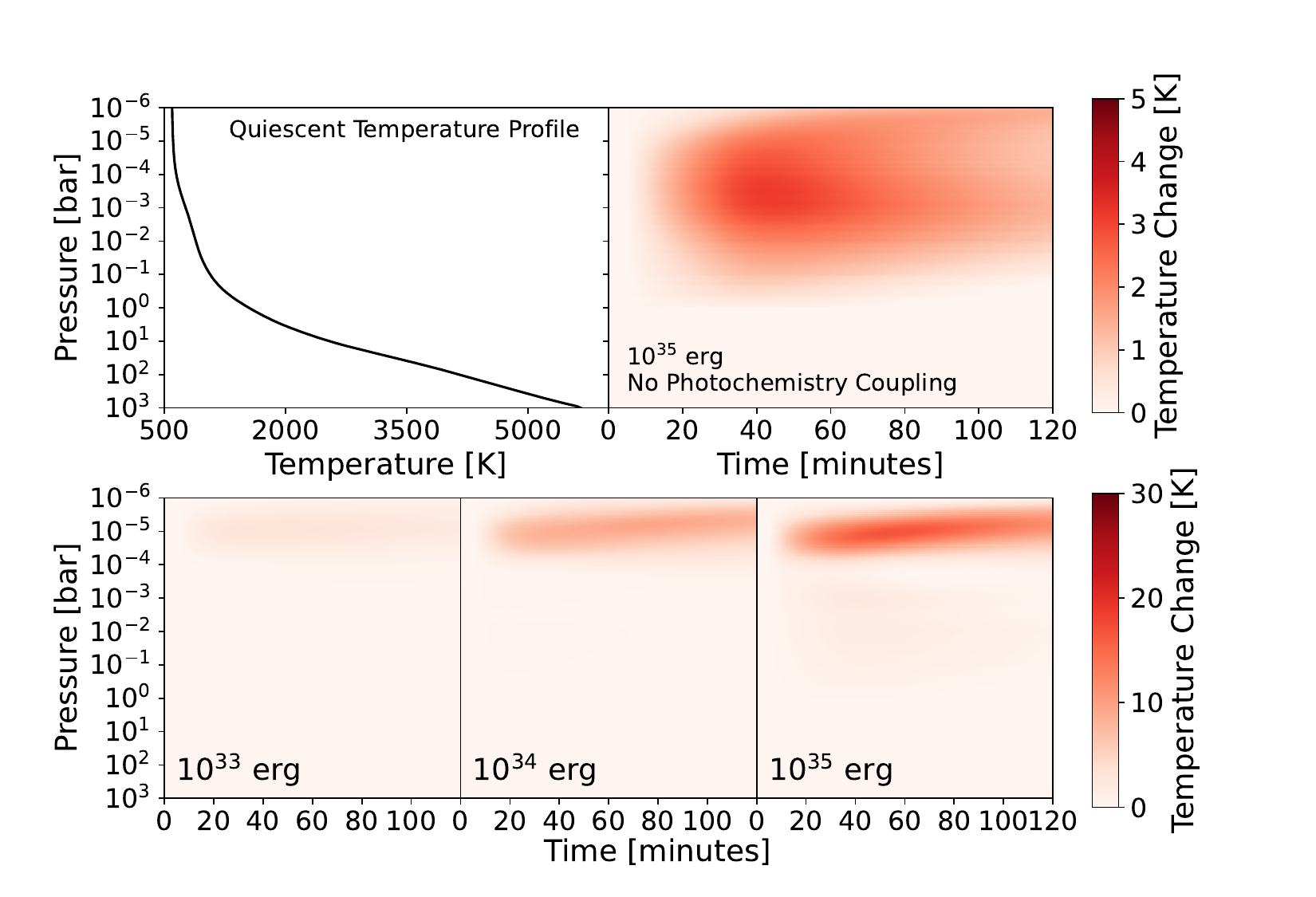}
\caption{\textbf{Change in temperature in response to flares of varying energy.} \textbf{Top left:} The initial temperature-pressure profile of the simulated atmosphere. \textbf{Top right:} The change in temperature in response to a $10^{35}\,$erg flare without considering any changes to atmospheric chemistry/opacity. Compared to simulations with temperature-chemistry coupling, the change in temperature is spread over a much broader pressure region of the atmosphere, resulting in a smaller temperature change. \textbf{Bottom:} Change in temperature with temperature-chemistry coupling included for flares of varying energy. The largest increase in temperature occurs in a narrow pressure region where most photochemistry is occurring, indicating that the change is caused primarily by a change in opacities. Negligible cooling $<1\,$K occurs anywhere in the atmosphere.}
\label{fig:temperature}
\end{figure*}

\begin{figure*}[t]
\centering
\includegraphics[width=1.0\textwidth]{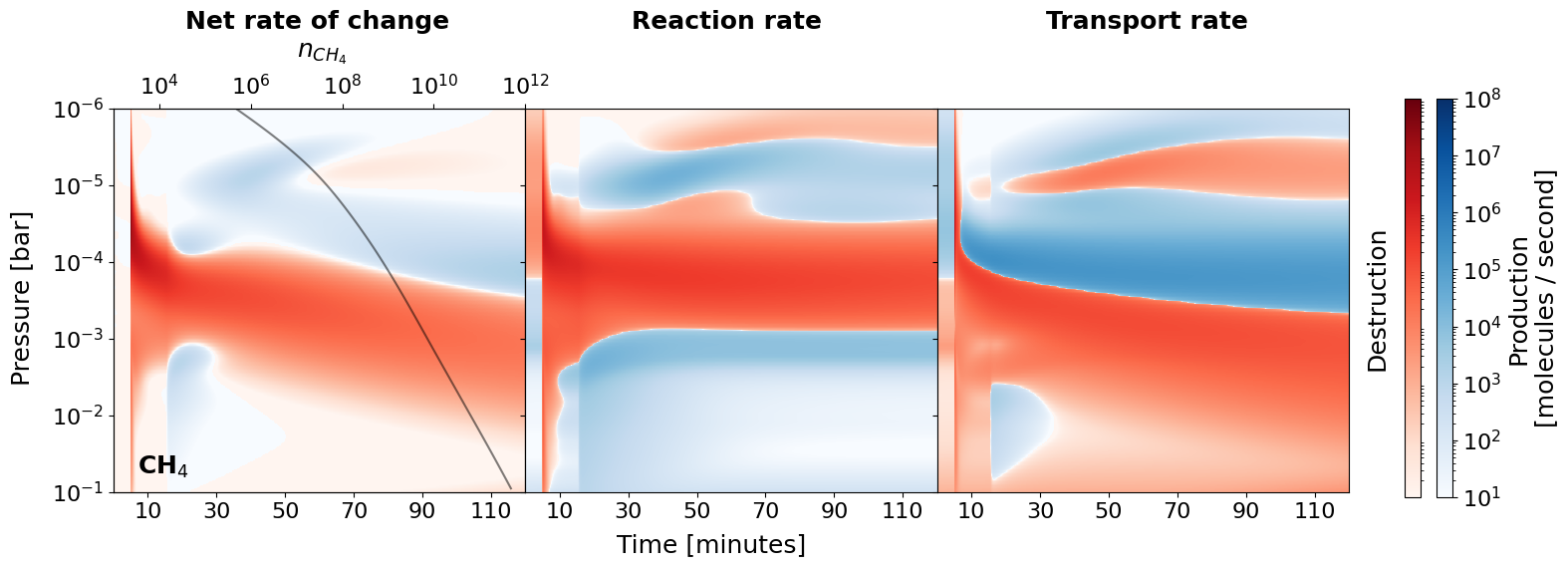}
\caption{\textbf{Contribution of reactions and transport to the net change in abundance of CH$_4$ in response to a $10^{34}\,$erg flare over two hours.} \textbf{Left:} The color plot shows the net rate of change in CH$_4$ number density at each timestep. Blue shows a net increase in CH$_4$ when considering both reactions and vertical transport (mixing), while red shows a net decrease. Both colors are plotted on the same absolute logarithmic scale. The initial CH$_4$ number density profile is plotted as a line for reference, corresponding to the top axis. Note that identical rates of change have a larger effect on the CH$_4$ mixing ratio at lower pressures due to reduced atmospheric density. \textbf{Center:} The summed chemical reaction rate affecting CH$_4$, across all reactions. Regions with matching colors in the left and center panels indicate levels where chemical reactions are the dominant mechanism for change in CH$_4$ abundance. \textbf{Right:} The rate of change of CH$_4$ due to vertical transport, calculated as the net rate minus the chemical rate. Regions of matching color in the left and right plots are levels where the change in CH$_4$ is primarily driven by mixing from other layers, rather than the chemical reaction rate.}
\label{fig:reaction_v_transport}
\end{figure*}

\begin{figure*}[t]
\centering
\includegraphics[width=1.0\textwidth]{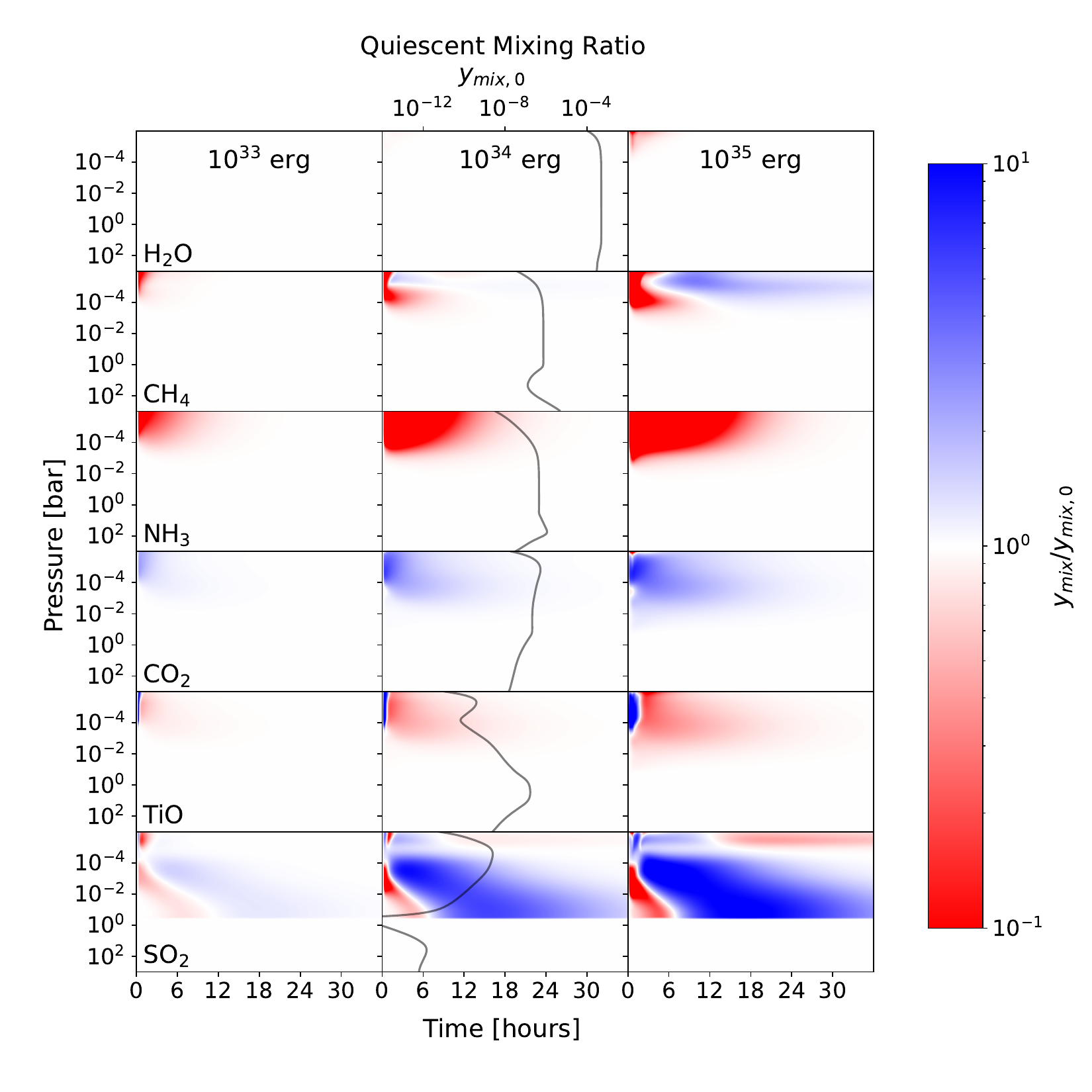}
\caption{\textbf{Photochemistry in response to single flares of varying energy.} No temperature-chemistry coupling is included in this simulation to reduce computational time. Initial chemical mixing profiles are shown as grey lines in the central column. More energetic flares have longer total durations, which drives differences in photochemistry behavior at early times for flares of different energies. }
\label{fig:36hour_chem}

\end{figure*}

Two species that remain relatively unaffected by the flare in terms of their relative mixing ratios are the dominant carbon- and oxygen-bearing species, \water\ and CO. CO's resistance to photochemistry stems from its extremely strong bond, the strongest among commonly occurring atmospheric species. \water, by contrast, is much more susceptible to photodissociation (primarily responsible for OH production). However, \water's high initial abundance means that, for a given photon flux, the change in relative mixing ratio is smaller when compared to trace species. As discussed further in Section \ref{sec:diss}, despite the small relative change in its mixing ratio, \water\ dissociation produces some of the most immediate and noticeable spectral changes in a flare-impacted brown dwarf.

Figure \ref{fig:1e34_2hour_chem} also illustrates how the type of photochemical response of certain species can change with time and pressure during and after the flare. Some notables examples are TiO and SO$_2$. TiO initially experiences an abundance enhancement but transitions to depletion around 40 minutes, corresponding roughly to the flare's end when the stellar flux returns to quiescent levels. While it is in general difficult to precisely track complex chemical pathways, some insights can be gained in the case of TiO, which is strongly tied to concentrations of TiO$_2$ and CO$_2$. In the first minute after the flare onset, the enhancement of TiO results primarily from the sudden and significant dissociation of TiO$_2$, which occurs faster than the dissociation of TiO. Despite the fact that the significant production of TiO by TiO$_2$ dissociation is very short-lived, shorter even than the duration of the flare, it leads to a temporary but significant buildup of TiO. As the flare continues and ends, however, the relative concentrations of TiO and TiO$_2$ continue to adjust to changes in other chemical species, eventually resulting in a destruction of TiO. Specifically, the reaction of TiO with CO$_2$, which has been photochemically enhanced, acts to convert TiO to TiO$_2$ at a greater rate than in the pre-flare state. This results in continued depletion of TiO until the CO$_2$ abundances also relax back to the pre-flare state. The importance of this pathway for TiO is supported by the nearly identical relaxation timescales of TiO and CO$_2$, shown in Figure \ref{fig:36hour_chem}, and TiO$_2$ also has a similar relaxation timescale.

The complex, pressure-dependent response of species like SO$_2$ is more challenging to interpret, as it depends on both photochemical and diffusive transport interactions within its chemical pathway. SO$_2$ is a known, observed photochemical tracer that is thought to be produced through an oxidation pathway of H$_2$S triggered by H and OH from water photolysis \citep{Hobbs2021_sulfur_chemistry,Tsai2023_WASP39_SO2}. Whether SO$_2$ is produced or destroyed depends on the abundance of H, OH, \water, and H$_2$S, among other species, and the photolysis rate of SO$_2$ itself. An exact description of why SO$_2$ is produced or destroyed in one layer versus another, and for how long, is beyond the scope of this paper, but the general trend of SO$_2$ production in most layers due to enhanced photochemistry from a flare is consistent with prior studies (e.g. \citealt{Tsai2023_WASP39_SO2}).   

One visualization of the interplay between reaction rates and mixing at different pressure levels is shown for CH$_4$ in Figure \ref{fig:reaction_v_transport}. This figure presents the net rate of change in number density of CH$_4$ and the separate contributions from chemical reaction and vertical transport rates. Several interesting features are visible. First, even in the quiescent state, the net reaction rate for a given species is rarely zero in any atmospheric layer, rather the equilibrium is created by transport between layers with chemical production and destruction. This highlights the complexity of atmospheric chemical pathways, as the most important reactions for a given species will not always occur in the same pressure region. Second, most of the destruction of CH$_4$ in response to the flare occurs extremely rapidly after the flare onset, within just the first few minutes. This trend is also seen for many other chemical species and shows how rapidly photodissociation reactions disrupt the quiescent steady state. Third, a marked change in chemical behavior occurs not only at the beginning of the flare, but also at the beginning of the exponential flare decay. In the case of CH$_4$, the region of net destruction between 1 and 0.1 mbar shrinks during the flare's decay. Finally, this plot demonstrates how the impact of reactions occurring in a narrow pressure region are mixed to a broader pressure region. In the case of CH$_4$, production following the flare occurs mostly around $10^{-5}\,$bar, however, the produced CH$_4$ is mixed into the surrounding atmospheric layers even though no direct chemical production is occurring there.

In addition to its chemical effects, a single flare also induces some thermal changes in the atmosphere. Figure \ref{fig:temperature} presents the temperature response to single flares. The top right panel shows the temperature response to a $10^{35}\,$erg flare in the absence of chemistry-temperature coupling. This simulation is performed with \texttt{HELIOS} only, with constant opacities throughout the run. The bottom right panel shows the response to the same flare both temperature and chemistry are coupled. Without coupling, the energy of the flare is absorbed across a broad pressure range from the top of the atmosphere to roughly $0.1\,$bar. The resulting temperature increase is very minimal--only a few Kelvin compared to the initial $\sim600-1000\,$K at those pressures--due to the large mass of gas being heated and the dominant influence of the brown dwarf's intrinsic heat over the external stellar flux.

When chemistry and temperature are coupled, the brown dwarf experiences larger, although still mostly insignificant heating of up to $\sim20\,$K, which is localized to a more narrow layer of the atmosphere from roughly $10^{-5}-10^{-4}\,$bar. The difference is caused by changing opacity in the upper atmosphere, and corresponds to the pressures with the most active photochemistry. Instead of directly absorbing more flare energy, this region becomes more opaque and retains both quiescent external and internal radiation more effectively, leading to sustained warming. This temperature increase persists beyond the flare duration and the layer's radiative cooling timescale, lasting as long as the underlying chemical changes that drive it.

\subsection{Impact of Flares of Different Energies}\label{sec:energies}

The photochemical impact of single flares of $10^{33}$, $10^{34}$, and $10^{35}\,$erg energies are shown in Figure \ref{fig:36hour_chem}. These simulations are run with \texttt{VULCAN} only to increase the simulation length to 36 hours and reduce computation time, meaning no temperature-chemistry coupling. The photochemical evolution observed in the first two hours of both the coupled and \texttt{VULCAN} only simulations are nearly identical. The small thermal impact of even the strongest flare means that leaving out temperature-chemistry coupling only changes the mixing ratios of any species by $<1\%$ in the first two hours, and likely by a similar amount across the entire simulation.

For most chemical species, the impact of a more energetic flare is to increase the magnitude of the same response seen at lower energies. For example, NH$_3$ and CO$_2$ are  destroyed and enhanced respectively for the smallest tested $10^{33}\,$erg flare, and more energetic flares simply increase the magnitude and length of the same response. Even species that experience more complex, pressure and time-dependent responses also generally only see an enhancement of the same pattern of behavior with energy. The complex behavior of SO$_2$ is enhanced by flares of higher energy, but the pressure and time behavior remains largely unchanged. Changes in the time-dependence are attributable to the length of the flare. Some behavior which appears emergent at higher flares energies, like the enhancement of CH$_4$ after several hours, is also occurring for lower energy flares, but is just at a very small scale not visible with the time and color scale of the plot.

The thermal response of the atmosphere to flares of different energies is shown in the bottom panels of Figure \ref{fig:temperature}. As expected, higher energy flares create larger temperature changes, but because the temperature impact is largely caused by photochemistry and opacity changes and not direct absorption of the flare's radiation, the temperature changes do not scale linearly with flare energy. A $10^{34}\,$erg flare causes about half the temperature change of a $10^{35}\,$erg flare.

\section{Discussion} \label{sec:diss}

\begin{figure*}[t]
\centering
\includegraphics[width=1.0\textwidth]{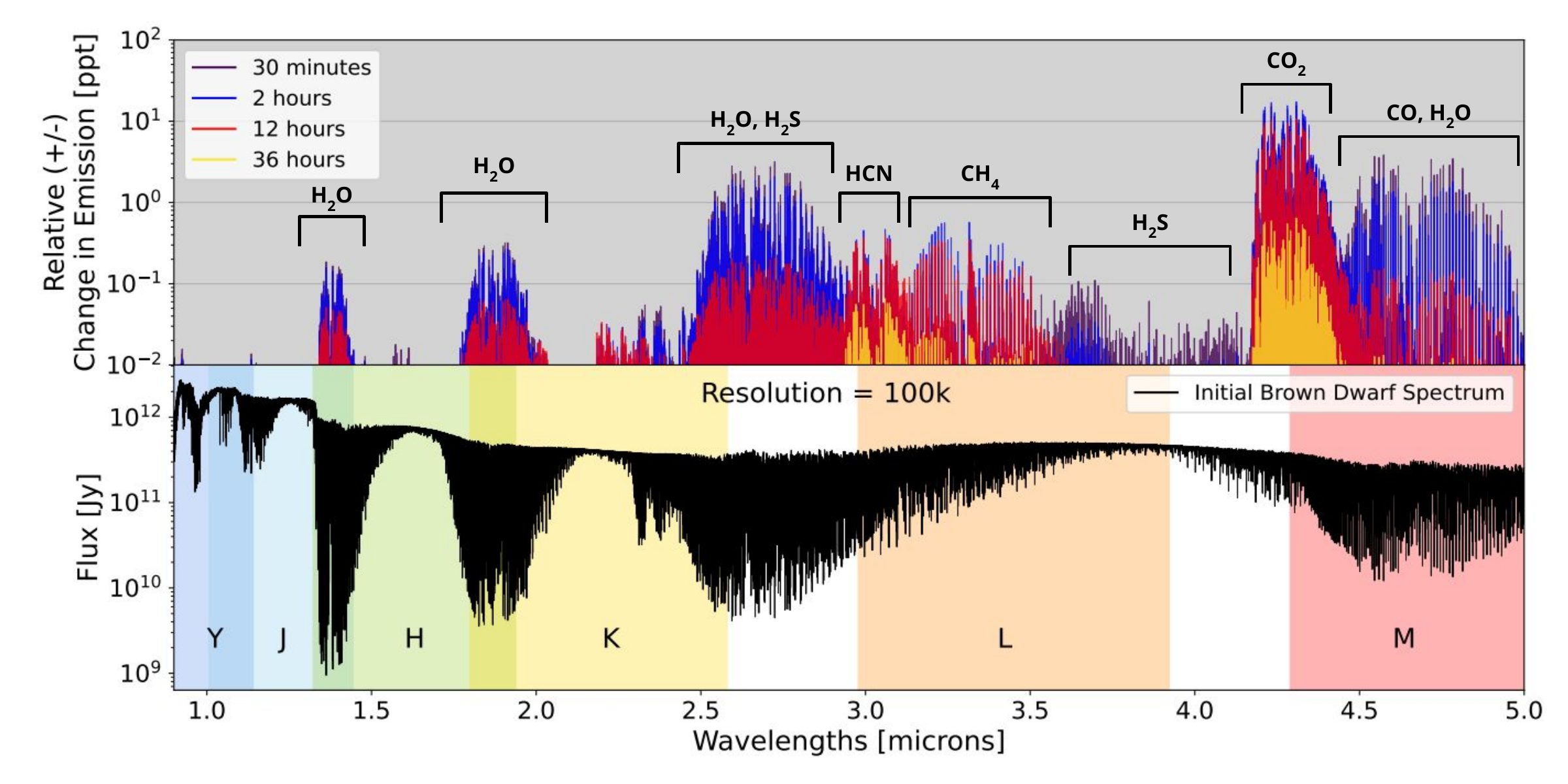}
\caption{\textbf{Relative change in the high-resolution (R$=10^5$) near-infrared emission spectrum of the brown dwarf in response to a single $10^{34}\,$erg flare.} \textbf{Top:} The relative change in the brown dwarf emission at four times spaced from the beginning of the simulated flare. All differences have been converted to positive values to allow plotting on a log-scale, though some changes correspond to flux increases and others correspond to flux decreases. Most change occurs due to dissociation of \water. At lower spectral resolution (R$=1000$, not shown), changes are significantly smaller, peaking at $\sim10\,$ppm, although the changes have a similar spectral and temporal pattern. Higher spectral resolution increases sensitivity to lower atmospheric pressures. \textbf{Bottom:} The initial brown dwarf emission spectrum, plotted with Earth's near-infrared observing windows.}
\label{fig:hr_spectrum}
\end{figure*}

\subsection{Comparison to Previous Studies} \label{sec:prev}

Flare-driven photochemistry has been explored in \htt\ dominated atmospheres previously by \citet{Konings2022}, \citet{Louca2023}, and \citet{Nicholls2023}. All of these studies have focused on simulating giant exoplanets, with lower surface gravities and temperatures than our simulation. Both \citet{Konings2022} and \citet{Louca2023} were limited to HCNO chemistry, whereas \citet{Nicholls2023} also includes chemistry of heavier metallic species like TiO and VO. \citet{Nicholls2023} has also investigated temperature-chemistry coupling, by solving for radiative-convective equilibrium at set time intervals as chemistry changes. \citet{Konings2022} has used a ``pseudo-2D'' framework to investigate horizontal transport of flare photochemistry from day to night sides.

Our findings largely replicate the results of \citet{Konings2022}, \citet{Louca2023}, and \citet{Nicholls2023} as applicable. For brevity, we will only discuss notable differences. All three papers note the significant photolysis of CH$_4$ in response to flares. This is also observed in the immediate aftermath of flares of any energy in our simulations, however, a few hours after the flare has ended, we instead see an increase in the abundance of CH$_4$ above the pre-flare state. This enhancement is a few orders-of-magnitude smaller than the initial depletion (essentially negligible for a $10^{33}\,$erg flare) and contained at lower pressure, but is nevertheless present. The exact chemical pathway for this CH$_4$ enhancement is difficult to pinpoint due to the high number of species involved in carbon chemistry and mixing between pressure levels, as discussed in Section \ref{sec:sfe}. From a initial overview of changing reaction rates, it appears that the enhancement of CH$_4$ is a result of introconversion between C$_2$H$_2$ and CH$_3$. C$_2$H$_2$ can itself be produced from reactions of CH$_4$, but it may be that another source, such as photolysis of CO, enhances the abundance of C$_2$H$_2$ that subsequently converts into CH$_4$ (see discussion of C$_2$H$_2$ photochemistry in \citealt{Moses2014_photochem_paper}). We note that this pathway is only our best guess, and other explanations are plausible. The reason that this CH$_4$ enhancement is not seen in other studies could be due to its relatively small magnitude (negligible for low energy flares), and it could also be an emergent feature of our higher temperature atmospheric regime. Our own tests of photochemistry in somewhat cooler brown dwarfs (T$_\textrm{eff}\sim1600\,$K), do not show this post-flare CH$_4$ enhancement, but are otherwise very similar.

We do not directly explore the impact of changing surface gravity on the atmospheric impact of flares, but can gain some insight from comparison to previous studies. For a given pressure, higher gravity atmospheres have less atmosphere at altitudes above them than a lower gravity atmosphere. This should mean, in general, that the impacts of photochemistry are felt at greater atmospheric pressures in high gravity atmospheres since the optical depth will be smaller, assuming that atmospheric opacities do not change significantly. Indeed, this has been shown before in simulations, for example by \citet{Baeyens2022_pseudo2d_photochem}. Comparing our results to \citet{Konings2022,Louca2023,Nicholls2023}, we infer the same pattern, where in our simulations, photochemical impacts for the same species are typically seen at one or two orders of magnitude higher pressures than in their lower gravity giant planet simulations.

In this work, we only explore the impact of single flares, while \citet{Konings2022}, \citet{Louca2023}, and \citet{Nicholls2023} also simulate multiple periodic or stochastic flares. Their analyses show that the impact of flares can build up over time to perturbed quasi-steady states. However, \citet{Konings2022} and \citet{Nicholls2023} showed that hotter atmospheres are faster to equilibrate because of their shorter chemical timescales. Our results bring into question whether flare impacts would significantly compound in a brown dwarf atmosphere. Even from an extremely powerful $10^{35}\,$erg flare, most chemical abundances have returned to their pre-flare state on the order of a day. Given enough time, we expect all chemical species to essentially return to their pre-flare abundances, likely on the order of a few days even for species with relatively slow reaction rates. The known active M dwarfs that host brown dwarfs only have flares of this energy on the order of every few weeks or months at most. More common are $10^{32}-10^{33}\,$erg optical flares, which can occur as often as every few days, but will have minimal atmospheric impact mostly lasting only half a day. If cumulative impacts do build up, it may only be in very select species, for example SO$_2$, which take a longer time to fully equilibrate even from lower energy flares. Of course, this entirely neglects the impact of particle radiation that may occur simultaneous to flares, which is likely to have delayed and longer-lasting effects. This also neglects any consideration of ionized species, which will certainly be an important part of the dynamics of upper atmosphere chemistry, but are not modeled in our framework.  

\subsection{Observability in High-resolution Emission Spectra} \label{sec:pRT_hrs}

One primary motivation for the investigation of the impact of stellar activity in brown dwarfs is their relative ease of observability compared to small exoplanets. To estimate the observability of the impact of single flares, we simulate the change in the emission spectrum of the brown dwarf at time intervals after a single $10^{34}\,$erg flare using the radiative-transfer code \texttt{petitRADTRANS} \citep{Molliere2019_pRT}. While many known brown dwarf companions are transiting their host, brown dwarfs are typically thought not amenable to transmission spectroscopy due to their high surface gravity which reduces their atmospheric scale heights to similar levels as Earth-like atmospheres. While emission spectroscopy, in contrast, is relatively easy for brown dwarfs, it creates a challenge in that most emission from high-gravity brown dwarfs is contributed by high pressure levels around $\sim1\textendash10$ bar, at altitudes below where flare impacts will be realized. This problem can be somewhat alleviated by observing with very high spectral resolution, which resolves individual line cores that are sensitive to much lower pressures, potentially lower than 1 mbar depending on resolution (e.g. \citealt{Xuan2022_bd_kpic}). We opt to use \texttt{petitRADTRANS} to generate our simulated emission spectra based because it can produce spectra up to R$=10^6$, while \texttt{HELIOS} is limited to lower resolution.

The top panel of figure \ref{fig:hr_spectrum} shows the change in emission in a simulated R$=10^5$ spectrum at near-infrared wavelengths where brown dwarfs are most luminous. A spectral resolution of around $10^5$ is around the highest resolution achieved by most current and planned ground-based near-infrared spectrographs. The change is presented in log-scale, meaning some changes are an increase in emission and some are a decrease. We only consider the change in chemical mixing ratios, and do not consider changes caused by temperature since they are not available beyond two hours. Temperature changes will enhance the changes in emission slightly, but at the small scale of the temperature change we observe in our simulations, it is by a small modification to the continuum flux rather than strongly modifying the relative strength of molecular absorption bands. 

In the immediate aftermath of a $10^{34}\,$erg flare, the spectrum can change by as much as a percent in certain wavelength regions. This drops off quickly after the end of the flare, however, with the difference in emission mostly less than 10 ppm after 36 hours. Most of the change in emission is related to the dissociation of water, which causes a decrease in opacity and increase in emission in between Earth's near-infrared observing windows, which are themselves caused by water opacity. Some less significant change in the \textit{K} and \textit{L} band is due to dissociation of H$_2$S, CH$_4$, and production of HCN. The most significant and long lasting change, however, is a decrease in emission caused by the production of CO$_2$ and occurs around $4.2\,$microns. Even after 36 hours, CO$_2$ is still responsible for a change in the emission spectrum of several hundred ppm. Unfortunately, this is not a spectral region accessible to most high-resolution spectrographs. All other chemical species, even with large changes in their mixing abundances, are not significant enough opacity sources to cause important spectral changes, at least from one flare. It may be possible that with repeated flaring, some species increase in spectral importance. For example, when compared to other species, HCN causes a relatively insignificant spectral change within 12 hours after the flare, but becomes more significant than everything except CO$_2$ by 36 hours, which could hint at better long-term visibility. Viewed at a lower spectral resolution of R$=1000$, the pattern of emission changes across all species remain similar, but with a much smaller magnitude, peaking around 10 ppm at most. This reinforces the importance of high-resolution to the observability of any flare impacts.

If the brown dwarf emission spectrum were to be directly observed without complications of a host star, current instruments likely could detect these flare impacts by cross-correlation methods if the brown dwarf was observed within a few hours of a major flare, as achieving signal-to-noise ratios above 100 are plausible with long integration times and cross-correlation methods can detect some changes below the noise level by leveraging many lines. However, what significance these changes would be detected at is unknown, and because these brown dwarfs are closely orbiting a M dwarf, any observation will have to contend with stellar spectral contamination. Stellar contamination, along with the short-time interval at which impacts will be visible, means that observing the impact of single flares is likely to be very challenging in brown dwarf atmospheres.

\section{Summary}\label{sec:conc}

Using a coupled chemical-kinetics and radiative-transfer framework, we simulate the impact of single energetic superflares on the atmosphere of a hot, high-gravity brown dwarf. Similar to the findings of \citet{Konings2022}, \citet{Louca2023}, and \citet{Nicholls2023} in hydrogen dominated exoplanet atmospheres, we find that flares can drastically change the mixing ratios of atmospheric gases in brown dwarfs, however, these changes are limited to low pressures at altitudes mostly above 1 mbar and short timescales of around a day. Despite large photochemical impacts, the thermal impact of even the most energetic flares is minimal (a few tens of Kelvin in our simulation), with most change in temperature resulting from photochemical changes in opacity rather than heating by the flare irradiation itself. This shows that in high-temperature, high-gravity brown dwarf atmospheres, self-consistent temperature-chemistry coupling is unnecessary when observability of flare impacts is the chief goal, and when only radiative heating from non-ionizing radiation is considered. However, future studies should investigate the importance of temperature-chemistry coupling when including heating from ionizing radiation and other non-radiative heating mechanisms. When photochemical changes are translated into changes in the emission spectrum of the brown dwarf at high spectral resolution, they can result in differences in emission of up to a percent in the immediate aftermath of the flare, but quickly fade away in the hours after the flare terminates. Due to fast chemical timescales of the high temperature brown dwarf, it is not clear that flares will occur frequently enough for significant accumulations of chemical and spectral changes over time. Despite the apparent difficulty of observing the impacts of flares in brown dwarf atmospheres, many unknowns remain, such as 3-D effects, differences due to charged ion chemistry, the impact of non-LTE thermodynamics in the high upper atmosphere, and the impact of particle radiation that may also occur alongside flares. Observing brown dwarfs that are subject to high levels of stellar activity is likely still worthwhile to validate models and explore these additional phenomena that are very difficult to simulate.

\begin{acknowledgments}

This work used computational and storage services associated with the Hoffman2 Cluster which is operated by the UCLA Office of Advanced Research Computing’s Research Technology Group.

The authors thank Brad Hansen, Hilke Schlichting, and Tuan Do for interesting discussions which helped to improve the paper.

\end{acknowledgments}


\facilities{UCLA Hoffman2 Cluster}


\software{Numpy \citep{vanderWalt2011}, Scipy \citep{Virtanen2019}, Astropy \citep{astropy:2018}}, VULCAN \citep{Tsai2017,Tsai2021}, HELIOS \citep{Malik2017,Malik2019}, petitRADTRANS \citep{Molliere2019_pRT}, Fiducial Flare \citep{Loyd2018}



\clearpage

\clearpage
\bibliography{references}
\bibliographystyle{aasjournal}



\end{document}